\documentclass[prl,superscriptaddress,showpacs,amssymb,amsmath,
amsfonts,aps,twocolumn,secnumarabic,floatfix]{revtex4}
\usepackage{graphics}
\usepackage{epsfig}
\begin{document} 
\bibliographystyle{try} 

\topmargin -0.9cm 
 
 \title{Unitarity constraints on Deeply Virtual Compton Scattering}

\newcommand*{\JLAB }{ Thomas Jefferson National Accelerator Facility, Newport News, Virginia 23606} 
\affiliation{\JLAB } 

\author{J.M.~Laget}
     \affiliation{\JLAB}

\date{\today} 

\begin{abstract} 

At moderately low momentum transfer ($-t$ up to 1 GeV$^2$) the coupling to the vector meson production channels gives the dominant contribution to real Compton and deeply virtual Compton scattering (DVCS). Starting from a Regge Pole approach that successfully describes vector meson production, the singular part of the corresponding box diagrams (where the intermediate vector meson-baryon pair propagates on-shell) is evaluated without any further assumptions (unitarity). Such a treatment explains not only the unexpectedly large DVCS unpolarized cross section that has been recently measured at Jefferson Laboratory (JLab), but also all the beam spin and charge asymmetries that has been measured at JLab and Hermes, without explicit need of Generalized Parton Distributions (GPD). The issue of the relationship between the two approaches is addressed. 

\end{abstract} 
 
 
\maketitle 

Deeply virtual Compton scattering (DVCS), $\gamma{^\star} p\rightarrow \gamma p$, is considered to be the process of choice to achieve an experimental determination of Generalized Parton Distributions (GPD). In the Bjorken regime (asymptotically large photon energy $\nu$ and virtuality $Q^2$, but fixed $x=Q^2/2m\nu$), the  amplitude factorizes into the coupling of the of incoming virtual photon ($\gamma^*$) and the outgoing real photon ($\gamma$) to a quark in the nucleon and its joint distribution in the initial and final states~\cite{Mu94,Ji97,Ra97}. In the Jefferson Laboratory (JLab) and Hermes at DESY energy ranges, current parmeterizations of the GPD lead to a DVCS amplitude much smaller  than the amplitude of the purely electromagnetic Bethe-Heitler (BH) process, where the real photon radiates from the incoming or the outgoing electron~\cite{Va99,Go01,Gui05}. This is the reason why it was proposed to access GPD through the determination of the interferences between the DVCS and BH amplitudes which enter the beam spin~\cite{Ai01,Ste01} and charge~\cite{Ai07,He02} or target~\cite{Ch06} asymmetries. Indeed such measurements are consistent with  expectations based on the factorization of the amplitude and simple models of GPD.  

\begin{figure}[hbt]
\begin{center}
\epsfig{file=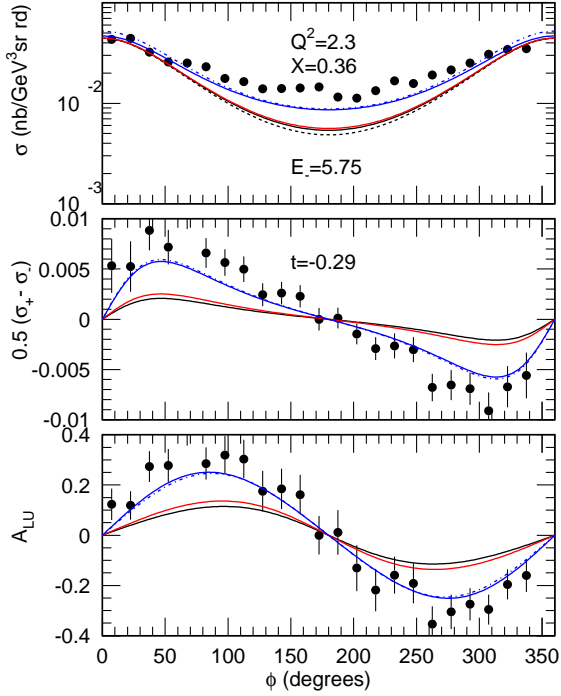,width=3.in}
\caption[]{(Color online) The Hall A DVCS cross sections and beam asymmetry are plotted against the azimuthal angle of the outgoing real photon. Top panel: unpolarized cross section, $d\sigma/dE_ed\Omega_edtd\phi$. Middle panel: Difference between the electron helicity dependent cross sections. Bottom panel: beam spin asymmetry. Black dashed curve: Bethe-Heitler contribution. Black solid curves: Regge pole contributions included. Red solid curves: coupling to $\rho$-nucleon channel included. Blue solid curves: coupling to diffractively produced intermediate states included. Blue dash-dotted: principal part of the rescattering integral included}
\label{HallA}
\end{center}
\end{figure}

The first measurement of the unpolarized cross section~\cite{Ca06} came as a surprise (Figure~\ref{HallA}). When plotted against the azimuthal angle $\phi$ of the emitted real photon (the angle between the electron scattering plane and the photon emission plane), the experimental cross section overwhelms the BH cross section by a factor three around $\phi=$~180$^{\circ}$.  The Compton cross section is smaller, by more than two orders of magnitude, than the meson production cross sections that dominate the total photo-absorption cross section: unitarity imposes a strong coupling between  these channels. This note provides a quantitative estimate of this effect that reproduces the available experimental observables, and addresses consequences on the determination of GPD.

\begin{figure}[hbt]
\begin{center}
\hspace{-1.5cm}
\epsfig{file=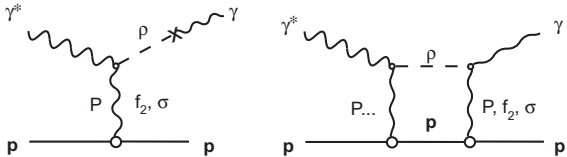,width=3.5in}
\caption[]{The relevant graphs in the $\gamma{^\star} p \rightarrow p \gamma$ reaction. Left: Poles + direct conversion. Right: $\rho$-nucleon unitary cut.}
\label{graph_cut}
\end{center}
\end{figure}

Figure~\ref{graph_cut} shows the two graphs that relate the Compton scattering amplitude and the $\rho$-nucleon channel. The first depicts the direct conversion of the vector meson into the real photon in the vacuum~\cite{Do01,Ca02,Ca03}. In the JLab or Hermes energy and momentum ranges, the energy of the emitted real photon is typically 2 to 4~GeV, and it fluctuates into a $\rho$ meson over a distance of 1.3 to 2.6~fm, larger than the size of the nucleon. In these conditions,  the Compton amplitude is proportional to the amplitude of the electro-production of a transversely polarized $\rho$ meson~\cite{Ca02,Ca03}:
\begin{eqnarray}
T_{\gamma^* \gamma}&=& T_{\gamma^*p\rightarrow \rho_{\perp} p}
\times\frac{\sqrt{4\pi\alpha_{em}}}{f_V}
\end{eqnarray}
where $\alpha_{em}$ is the fine structure constant and $f_V$ is the radiative decay constant of the vector meson. I keep only the photo-production of the $\rho$, since it dominates over the production of $\omega$ and $\phi$. I use the latest version~\cite{La04} of the Regge pole exchange model~\cite{Ca02,Ca03} which reproduces the experimental cross section of $\rho$ photo-production~\cite{BaXX}, as well as the cross section of the electro-production of a transversely polarized $\rho$~\cite{CoXX,HeXX,CyXX,Gui07} in the JLab/Hermes energy range. 

The Compton amplitude is projected onto the helicities of the incoming and outgoing particles, combined with the electromagnetic tensor and  added coherently to the purely electromagnetic Bethe-Heitler amplitude.  While it accounts for about one third of the beam spin asymmetry, it is not strong enough to reproduce the unpolarized experimental cross section and the difference between the polarized cross sections.

The second graph in Fig.~\ref{graph_cut} depicts the conversion of the vector meson into the real photon in the field of the nucleon. 
The corresponding rescattering amplitude takes the form:
\begin{eqnarray}  
T_{\gamma^* \gamma}&=& \int \frac{d^3\vec{p}}{(2\pi)^3}
\frac{m}{E_p}
 \frac{1}{P^2_{\rho}-m^2_{\rho}+i\epsilon}
%\nonumber\\ &&  \times 
T_{\gamma^*p\rightarrow \rho p}T_{\rho p \rightarrow \gamma p}
\end{eqnarray}
where the integral runs over the three momentum $\vec{p}$ of the intermediate nucleon, of which the mass is $m$ and the energy is $E_p=\sqrt{p^2+m^2}$. The four momentum  and the mass of the intermediate $\rho$ are respectively $P_{\rho}$ and $m_{\rho}$. The integral can be split into a singular part, that involves on-shell matrix elements, and a principal part~$\cal{P}$:
\begin{eqnarray}
 T_{\gamma^* \gamma}&=& -i\frac{p_{c.m.}}{16\pi^2}\frac{m}{\sqrt{s}}
 \int  d{\Omega} \left[T_{\gamma^*p\rightarrow \rho p}(t_{\gamma^*})
 T_{\rho p \rightarrow \gamma p}(t_{\gamma})
 \right]
 \nonumber \\ && 
 +\cal{P}
 \label{sing}
\end{eqnarray}
where $p_{c.m.}=\sqrt{(s-(m^2_{\rho}-m^2))(s-(m^2_{\rho}+m^2))/4s}$  is the on-shell momentum  of the intermediate proton, for  the c.m. energy $\sqrt s$. The two fold integral runs over the solid angle $\Omega$ of the intermediate proton. The four momentum transfer between the incoming virtual photon and the $\rho$ is $t_{\gamma^*}=(k_{\gamma^*}-P_{\rho})^2$, while the four momentum transfer between the $\rho$ and the outgoing real photon is $t_{\gamma}=(k_{\gamma}-P_{\rho})^2$. The summation over all the spin indices of the intermediate particles is meant. However since vector meson photo and electro-production conserve helicity, this sum is trivial and the intermediate vector meson is transversally polarized.

The singular part of the integral relies entirely on on-shell matrix elements for photo-production~\cite{La00} and electro-production~\cite{Ca03,La04} of vector mesons which reproduce the world set of data. At low $-t$ (up to 1~GeV$^2$),  they are mostly diffractive and can be expressed in the form:
\begin{eqnarray}
T_{\gamma^*p\rightarrow \rho p}(t_{\gamma^*}) &=&
 T_{\gamma^*p\rightarrow \rho p}(t)\exp(\frac{b^*}{2}(t_{\gamma^*}-t)
\end{eqnarray}
\begin{eqnarray}
T_{ \rho p\rightarrow \gamma p}(t_{\gamma})&=&i | T_{\gamma p\rightarrow \rho p}(0)| \exp(\frac{b}{2}t_{\gamma})
\end{eqnarray}
where $t=(k_{\gamma}-k_{\gamma^*})^2$ is the four momentum transfer between the incoming virtual and the outgoing real photons. This allows to factorize the $\rho$ electro-production elementary amplitude~\cite{La04} and to combine the pole and the rescattering amplitudes in a compact form:
\begin{eqnarray}
T_{\gamma^* \gamma}&=&  T_{\gamma^*p\rightarrow \rho_{\perp} p}
\left[ \frac{\sqrt{4\pi\alpha_{em}}}{f_V} +  
\frac{p_{c.m.}}{16\pi^2}\frac{|T_{\rho p\rightarrow \gamma p}(0)|}{\exp(\frac{b^*}{2}t)} 
\right . \nonumber \\
&&\left . \times (1+iR) \frac{m}{\sqrt{s}}\int d\Omega_p \exp(\frac{b^*}{2}t_{\gamma^*})\exp(\frac{b}{2}t_{\gamma})
\right]
\label{amp}
\end{eqnarray}
where $R$ stands for the ratio between the principal and the singular parts of the rescattering integral. The forward photo-production amplitude can be related to the experimental cross section as follows:
\begin{eqnarray}
|T_{\rho p\rightarrow \gamma p}(0)|&=&(s-m^2)\frac{2\sqrt \pi}{m} \sqrt{\frac{d\sigma}{dt}(0)}
\end{eqnarray}

Using $4\pi/f^2_V= 0.494$ ~\cite{Do01,Ca03}, $\frac{d\sigma}{dt}(0)$ =130~$\mu$b/GeV$^2$ (Fig.~5.6 of~\cite{LaXX}), $b=$ 6~GeV$^{-2}$ and $b^*=$ 2 GeV$^{-2}$ (fit of the slope of the photo- and electro-production of $\rho$ below $-t=$ 1 GeV$^2$), one gets in the kinematics of Fig.~\ref{HallA} ($Q^2=$ 2.3~GeV$^2$, $x=$ 0.36, $\sqrt s =$ 2.269~GeV and $t=$ -0.29 GeV$^2$):
\begin{eqnarray}
T_{\gamma^* \gamma}&=& T_{\gamma^*p\rightarrow \rho_{\perp} p}
\left[ 0.060 + 0.018(1+iR) \right]
\label{cut1}
\end{eqnarray}

The contribution of the rescattering integral is comparable to the Regge pole amplitude. Theoretical predictions move toward experimental observables, but the unpolarized cross section is still underestimated around $\phi=$ 180$^{\circ}$. 
Since the cross section of the elastic photo-production of $\rho$ (namely the $\gamma p\rightarrow \rho p$ channel) amounts only to 15~\% of the total photo-absorption cross section, we have also to couple the Compton scattering amplitude to the inelastic channels that build up the total absorption cross section. While it is an almost impossible task to compute each individual inelastic channel, it is possible to get an estimate of the sum of their contributions to the unitary rescattering integral. In the JLab and Hermes energy ranges only diffractive inelastic channels ({\it e.g.} $\gamma p \rightarrow {\rho^*} p$, $\gamma p \rightarrow \rho$N$^*$, \ldots) survive, and it is reasonable to assume that their production amplitude has the same structure as the amplitude of the production of a $\rho p$ pair. Indeed, the exchange of the Pomeron and the $f_2$ reproduces the total photo absorption $\sigma_{abs}$ (Fig.~5.2 of~\cite{LaXX}), as well as the elastic $\rho$ production cross section $\sigma_{\rho}$~\cite{La00}.   Under this assumption the rescattering integral in eq.~\ref{cut1} is simply renormalized by the ratio $\sigma_{abs}/\sigma_{\rho}$):
\begin{eqnarray}
T_{\gamma^* \gamma}&=& T_{\gamma^*p\rightarrow \rho_{\perp} p}
\left[ 0.060 + 0.018\frac{\sigma_{abs}}{\sigma_{\rho}} (1+iR)\right]
\label{cut2}
\end{eqnarray}

\begin{figure}[hbt]
\begin{center}
\epsfig{file=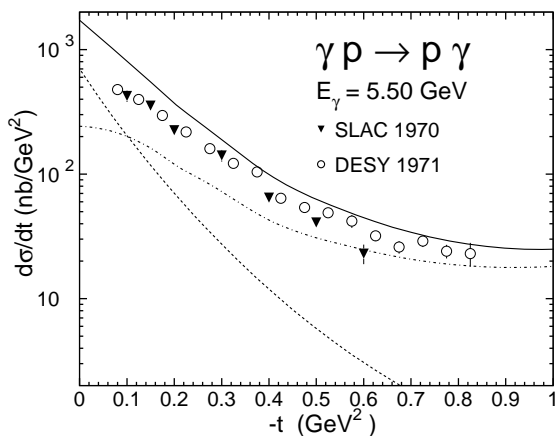,width=3.in}
\caption[]{The real Compton scattering cross section at $\sqrt s =$ 3.35~GeV. Dashed curve: Pole terms. Dot-dashed curve: rescattering only. Full curve: both contributions.}
\label{real}
\end{center}
\end{figure}

This procedure leads to a good agreement with the real Compton scattering cross section~\cite{An70,Bu71} at moderate $-t$ (up to 1~GeV$^2$) as depicted in Fig.~\ref{real}. In this case, $\sqrt s =$ 3.35~GeV, $b^*=b=$ 7~GeV$^{-2}$,  $\frac{d\sigma}{dt}(0)$ =100~$\mu$b/GeV$^2$,  and eq.~\ref{cut2} reads, at $-t=$ 0.15~GeV$^2$:
\begin{eqnarray}
T_{\gamma \gamma}&=& T_{\gamma p\rightarrow \rho_{\perp} p}
\left[ 0.060 + 0.013\frac{\sigma_{abs}}{\sigma_{\rho}} (1+iR) \right]
\label{cut3}
\end{eqnarray}
with $\sigma_{abs}/\sigma_{\rho}=$ 6, which lies in the lower part of the range of the experimental ratio  $\sigma_{abs}/\sigma_{\rho}= (125\pm10) /(15\pm 2 )=$ 7.8~$\pm$~1.6 (Figs.~29 and 76 of~\cite{Ba78}). 

\begin{figure}[hbt]
\begin{center}
\epsfig{file=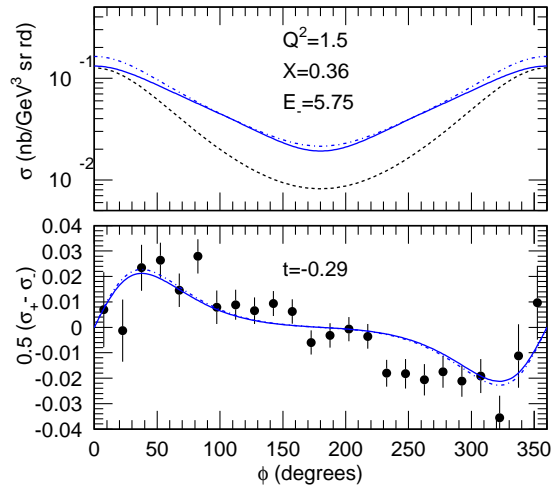,width=3.in}
\caption[]{The same as in Fig.~\ref{HallA}, but for Q$^2$= 1.5~GeV$^2$.  }
\label{HallA_15}
\end{center}
\end{figure}

For virtual photons, $\sigma_{abs}/\sigma_{\rho}= (13.70\pm0.24) /(1.28\pm 0.37 )=$ 10.70~$\pm$~3.24, at Q$^2$= 2.3~GeV$^2$  and $x=0.36$, from the SLAC~\cite{Go94}, JLab~\cite{CyXX,Gui07} and Cornel~\cite{CoXX} data. The contribution of the singular part of the inelastic cuts accounts for the major part of the JLab cross sections in Fig.~\ref{HallA} at Q$^2$= 2.3~GeV$^2$, as well at Q$^2$= 1.5~GeV$^2$ in Fig.~\ref{HallA_15}, with $\sigma_{abs}/\sigma_{\rho}=$ 8.3. This value lies in the lower part of the range of the experimental value, consistently with the analysis of the real photon sector (Fig.~\ref{real}). The choice of the central experimental values (10.7) increases by 17\% the cross section at $\phi=$ 180$^{\circ}$, as well as their difference at 90$^{\circ}$. Since the evaluation of the principal part requires extra assumptions on the off-shell behavior  of the elementary meson production amplitudes, I postpone it to a future publication. In order to get an estimate of its size, I note that in real Compton scattering at forward angles, $R$ ranges from -0.5 at $\sqrt{s}=$ 2.2~GeV to -0.23 at $\sqrt{s}=$ 3.35~GeV~\cite{Ba78}. The dash-dotted curves in Figs.~\ref{HallA} and~\ref{HallA_15} have been evaluated with $R=-0.5$. 

\begin{figure}[hbt]
\begin{center}
\epsfig{file=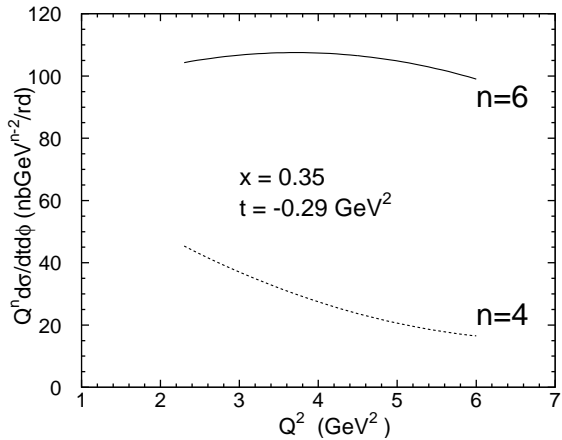,width=3.in}
\caption[]{The evolution with $Q^2$ of the differential cross section of the hadronic DVCS alone, at $x$ =~0.35 and $t$=~-0.29 Gev$^2$.  The dashed curve is scaled by Q$^4$. The solid curve is scaled by Q$^6$. }
\label{scaling}
\end{center}
\end{figure}

The model reproduces the evolution with Q$^2$ of the polarized cross section difference in the limited range that has been covered by the JLab/hall A experiment. Fig.~\ref{scaling} shows the expected scaling behavior of the hadronic DVCS cross section alone (without the BH contribution) up to the highest photon virtuality that is accessible with a beam of $E_-$=~11 GeV at $x$=~0.35. It decreases as $1/Q^6$, faster than the leading twist cross section based on GPD ($1/Q^4$).

\begin{figure}[hbt]
\begin{center}
\epsfig{file=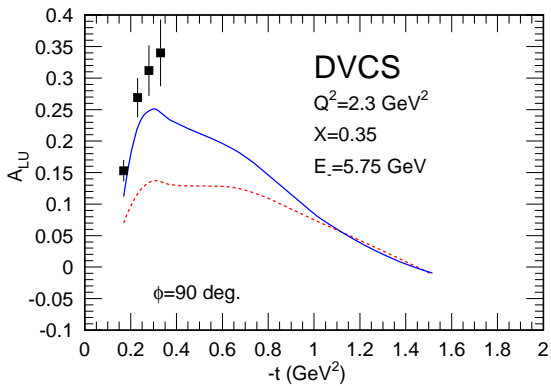,width=3.in}
\caption[]{(Color online) The $t$ distribution of the beam spin asymmetry at $\phi$ =~90$^{\circ}$, at JLab. Squares: Hall~A~\cite{Ca06}. The red (dashed) curve corresponds to the elastic $\rho p$  intermediate state. The blue solid curve corresponds to all the diffractive inelastic states. }
\label{A_LU_t}
\end{center}
\end{figure}

The model also agrees fairly well with the $t$ dependency of the beam spin asymmetry $A_{LU}$  that is deduced from the recent JLab Hall A cross sections~\cite{Ca06} and predicts its evolution at higher $-t$ in Fig.~\ref{A_LU_t}.  It will be compared to the Hall~B full data set in the experimental paper~\cite{FX07}: suffice to say that it leads to a fair account of the magnitude and the dependency of the experimental beam spin asymmetry in a wide range ($1.5<Q^2<3.5$~GeV$^2$, $0.15<x<0.45$, $-t$ up to 1.2~GeV$^2$).

\begin{figure}[hbt]
\begin{center}
\epsfig{file=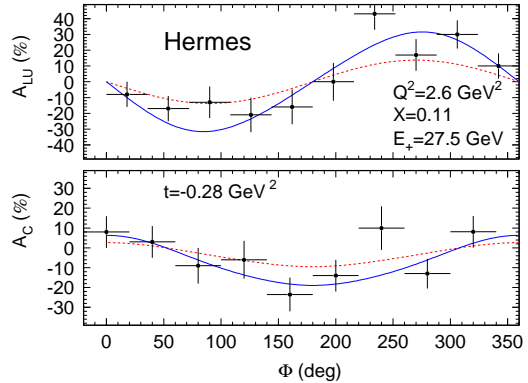,width=3.in}
\caption[]{(Color online) The azimuthal distribution of the beam spin (top) and charge (bottom) asymmetries at Hermes. The red (dashed) curves correspond to the elastic $\rho p$  intermediate state. The blue solid curves correspond to all the diffractive inelastic states. }
\label{hermes}
\end{center}
\end{figure}

Finally the beam spin~\cite{Ai01} and charge~\cite{Ai07,He02} asymmetries that have been measured at Hermes are also well accounted for (Figure~\ref{hermes}).

In summary, I suggest that the coupling with vector meson production channels represents the major part of the DVCS amplitude at reasonably low $Q^2$ (up to about $3$~GeV$^2$). The coupling to the elastic $\rho p$ channel is on solid ground, since it relies on on-shell matrix elements that are driven by $\rho$ photo- and electro-production  measured cross sections. It provides us with a reliable lower limit of the DVCS cross sections and observables. The coupling with the inelastic channels leads to a fair account of all the DVCS observables that have been recorded so far. This conjecture has to be kept in mind in any attempt to access and determine the GPD at the low virtuality $Q^2$ currently available at JLab or Hermes. On the one hand, the available energy is too low ($\sqrt{s}<$ 2.5~GeV at JLab) to sum over the complete basis of hadronic intermediate states, and safely rely on a dual description at the partonic level. On the other hand, unitarity  relates the Compton scattering amplitude to the transverse part of the vector meson production amplitudes. While the DVCS and the longitudinal part of the light meson production amplitudes factorizes~\cite{Co99,Co97} at the leading order in $1/Q$ into the quark current and GPD, the transverse part of meson electro-production does not. Therefore the implementation of the unitarity constraint at the hadronic level provides us with a measure of the size of higher order contributions in DVCS.

At higher $Q^2$, the available energy increases (at fixed $x$) and a partonic description may become more justified: Above which $Q^2$ is an open issue. Also,  the relative importance  of the contribution of the coupling to vector meson channels and the factorized amplitude is another open issue. It depends on how fast the transverse cross section of vector meson production decreases with $Q^2$. Although it is encouraging that the hadronic DVCS cross section scales as $1/Q^6$ and decreases faster than the factorized hard cross section ($1/Q^4$), any sensible program, that is aimed at determining the GPD with the DVCS reaction, cannot make the economy of a concurrent determination of the transverse and longitudinal cross sections of the elastic and inelastic diffractive production of vector mesons.

I acknowledge discussions with P. Bertin and C. Hyde-Wright whose the experiment triggered this work, A. Radyushkin and M. Vanderhaeghen. I acknowledge the warm hospitality at JLab where this work was completed. Jefferson Science Associates operates Thomas Jefferson National Facility for the United States Department of Energy under contract DE-AC05-06OR23177.

\end{document}